\documentclass[12pt]{article}
\input epsf
\title{\bf Single and double inclusive cross-sections
for nucleus-nucleus collisions in the perturbative QCD}
\author{M.A.Braun\\
Dep. of High Energy physics,
 University of S.Petersburg,\\
198504 S.Petersburg, Russia   }
\def\beq{\begin{equation}}
\def\eeq{\end{equation}}
\def\noi{\noindent}
\def\psid{\psi^{\dagger}}
\def\tchi{\tilde{\chi}}

\begin{document}
\maketitle
\medskip
\noi{\bf Abstract.}
Single and double inclusive cross-sections in nucleus-nucleus
collisions are derived in the perturbative QCD with interacting
BFKL pomerons in the quasi-classical approximation.

\section{Introduction}
With the advent of colliders the study of particle production in
nucleus-nucleus collisions has acquired a prominent role both in
experimental and theoretical studies. Single and double inclusive
cross-sections and correlations related to them  draw naturally most
attention. In the framework of the perturbative QCD they can be
studied in the approach based on interacting BFKL pomerons
developed for nucleus-nucleus collisions in ~\cite{bra1,bra2,bra3}.
Intuitive considerations plus experience with the Local Reggeon Field
Theory (LRFT) allow to guess the structure of the single inclusive
cross-section as emission either from the central pomeron or from the two
adjacent triple pomeron vertexes in the convolution of two sets of fan
diagrams propagating from the center to the projectile or target.
However no rigorous demonstration of this structure has been given
in the literature. Still worse is the situation with the double
inclusive cross-section for which even in the LRFT one has a very
complicated formula (see ~\cite{CM}).

In this paper we aim at filling this gap. We derive formulas for both
single and double inclusive cross-sections in
nucleus-nucleus collisions in the perturbative QCD approach with
interacting BFKL pomerons. Part of the problem which involves particle
emission from the pomerons is ideologically rather similar to the
LRFT (although considerably more complicated technically). So to treat
this problem we shall use the cut pomeron formalism developed within
the LRFT ~\cite{ital}, which we appropriately generalize for the BFKL pomerons.
A new problem is emission from the triple pomeron vertex.
For the single inclusive we find  that it is
indeed described by simple formulas used in our previous
studies (~\cite{bra4}). However the double inclusive cross-sections
with a single emission from the vertex lead to a complicated
expression including
different parts of the cut emission vertex found in ~\cite{bra5}.

Numerical calculations of the single inclusive cross-section in
nucleus-nucleus collisions, with
vertex emission taken into account,
were earlier reported in ~\cite{bra4} (as mentioned, without rigorous
justification of the formulas). As to the double inclusive
cross-sections and correlations, due to their complexity, we do not
attempt here to use them for numerical studies.
These studies present a separate difficult problem and are postponed for
future publications.

The paper is organized as follows. In the next  section we generalize
the cut pomeron formalism for BFKL pomerons. The derivation closely
follows that in the LRFT (~\cite{CM,ital}) with inevitable complications
due to the non-local structure of the BFKL pomeron. Next in Section 3
we solve the equation of motion and find working formulas for the
single and double inclusive cross-section corresponding to emission
from the pomeron. In Section 4 we discuss emission from the vertex
and derive formulas for the single and double inclusive cross-sections
which involve such emission. Finally we discuss our results
in the  last section.

\section{Cut pomeron formalism and inclusive cross-sections
for nucleus-nucleus scattering
in the perturbative QCD}
\subsection{Fields and Lagrangian}
Pomerons are described by two  fields $\psi$ and $\psid$, which depend
on rapidity $y$, relative gluon momentum $q$ and point $b$ in the
transverse space.
We shall restrict ourselves with the purely forward case relevant
for collisions off large nuclei. Then  $b$ is conserved during the
collision. Correspondingly we do not indicate it explicitly
unless it may lead to confusion.
After transition to real fields the generating functional for the
Green functions is written as
\beq
Z=\int D\psi D\psid e^{-A},
\eeq
where action $A$ is an integral over $y$, $q$ and $b$:
\beq
-A=\int dy d^2bd^2qL(\psi,\psid)
\eeq
and the Lagrangian density $L$ is
\beq
L(y,q,b)=\psid Q\psi+\lambda\Big[(K\psid)\cdot\psi^2+(K\psi)\cdot
{\psid}^2\Big]+g\delta(y)\psi+f\delta(Y-y)\psid.
\eeq
Here $Q$ is essentially the BFKL operator
\beq
Q=2K\Big(\frac{\partial}{\partial y}+H\Big),
\eeq
with $H$ the BFKL Hamiltonian, and
\beq
K=\nabla_q^2q^4\nabla_q^2.
\eeq
Operator $K$ is conformal invariant and commutes with $H$.
The triple-pomeron coupling $\lambda$ is
\beq
\lambda=\frac{4\alpha_s^2N_c}{\pi}.
\eeq
Finally $g(y,b)$ and $f(y,b,B)$ represent the coupling of the pomeron to
the colliding nuclei. $g(y,b)$ describes this coupling to the target A
(at zero rapidity)
\beq
g(y,b)=AT_A(b)\rho(q),
\eeq
where $T_A(b)$ is the profile function of nucleus A with its center at the
origin in the transverse plane
and $\rho(q)$ is the colour density of the nucleon. $f(y,b,B)$ describes
the coupling to the
projectile at rapidity $Y$ and its center at point $B$ (the impact
parameter of the collision)
\beq
f(y,b,B)=BT_B(B-b)\rho(q).
\eeq

To study different cuts we introduce the cut pomeron
formalism following the scheme developed in ~\cite{ital} for the local
pomeron. We introduce 6 fields
\beq
\psi_{\pm},\ \ \psi_{c},\ \ \psid_{\pm},\ \ \psid_c
\eeq
which describe pomerons to the left (right) of the cut
(subindex +(-)) and cut pomerons (subindex $c$).
To fulfill the AGK rules the Lagrangian of these fields is to be taken as
\beq
L_c=L_c^0+L_c^I+L_c^E.
\eeq
Here the unperturbed Lagrangian is
\beq
L_c^0=\sum_{i=\pm,c}\epsilon_i\psid_iQ\psi_i,
\eeq
with
\[\epsilon_{\pm}=-\epsilon_c=1.\]
The interaction involves several terms
\[
L_c^I=\lambda(K\psid_+)\cdot{\psi}_+^2
+\lambda(K\psi_+)\cdot{\psid_+}^2
+\lambda(K\psid_-)\cdot{\psi}_-^2
+\lambda(K\psi_-)\cdot{\psid_-}^2\]\[
-2\lambda (K\psid_c)\cdot\psi_c(\psi_++\psi_-)
-2\lambda (K\psi_c)\cdot\psid_c(\psid_++\psid_-)\]\beq
+\sqrt{2}\lambda (K\psid_c)\cdot(\psi_+\psi_-+\psi_c^2)
+\sqrt{2}\lambda(K\psi_c)\cdot(\psid_+\psid_-+{\psid_c}^2).
\label{inter}
\eeq
Finally the external Lagrangian is
\beq
L_c^E=g\delta(y)(\psi_++\psi_--\sqrt{2}\psi_c)+
f\delta(Y-y)(\psid_++\psid_--\sqrt{2}\psid_c).
\eeq

\subsection{Inclusive cross-sections}
To study  inclusive cross-sections we additionally introduce
interaction of the cut pomeron with the emitted particles.
As is well-known, the inclusive cross-section corresponding to emission
of a gluon jet from the pomeron can be written as a double integral
\beq
I(y,\kappa)\equiv \frac{(2\pi)^3d\sigma}{dyd^2\kappa}=2\int d^2b
d^2q_1d^2q_2\delta^2(q_1+q_2-\kappa)\eta(q_1,q_2)p_A(y,q_1,b)p_B(y,q_2,b),
\eeq
where $p_{A(B)}(y,q,b)$ is  the pomeron coupled to the target (projectile)
and
\beq
\eta(q_1,q_2)=\frac{16\pi^2\alpha_sN_c}{\kappa^2}q_1^2\nabla_1^2q_2^2\nabla_2^2.
\label{eta}
\eeq
This expression can be conveniently rewritten in the coordinate space in the local form,
in terms of the 'non-amputated' pomeron
\beq
P(y,r,b)=2\pi r^2\int \frac{d^2q}{2\pi}e^{iqr}p(y,q,b)
\label{fromqtor}
\eeq
as
\beq
I(y,\kappa)=\frac{8 \alpha_s N_c}{\kappa^2}\int d^2bd^2r P_A(y,r,b)\nabla^2 e^{i\kappa r}
\nabla^2P_B(y,r,b).
\eeq

To generate inclusive cross-sections we add a term to the action
\beq
-\Delta A_c=\int
dyd^2bd^2q_1d^2q_2\xi(y,q_1,q_2,b)\psid_c(y,q_1,b)\psi_c(y,q_2,b).
\label{delta}
\eeq
Differentiation in $\xi$ then gives insertions into the cut
pomeron propagator
at rapidity $y$ and transverse point $b$ with momenta $q_1$ and $q_2$.
As in ~\cite{CM} we denote the total action as
\beq
A_c(\xi)=A_c+\Delta A_c.
\eeq
%%%%%%%%%%%%%%%%%%%%%%%%%%%
It follows from unitarity that
\beq
A_c(\xi=0)=0
\label{zero}
\eeq
Also one finds
that the generating functional of the amplitudes at
$\xi\neq 0$ is given by
%%%%%%%%%%%%%%%%%%%%%%%%%%
\beq
T_c(\xi)=1-S+\frac{1}{2}\Big(e^{-A_c(\xi)}-1\Big)
\label{tcxi}
\eeq
where $S$ is just the $S$ matrix at $\xi=0$  ~\cite{CM}  (also
see Appendix 1).

The single and double inclusive cross-sections are obtained as
\beq
I_1(y,\kappa)=2\int d^2b d^2q_1d^2q_2\delta^2(q_1+q_2-\kappa)\eta(q_1,q_2)
\frac{\delta
T_c(\xi)}{\delta\xi(y,q_1,q_2,b)}\Big|_{\xi=0}
\label{single0}
\eeq
and
\[
I_2(y,\kappa|y',\kappa')=2\int d^2b d^2b'
d^2q_1d^2q_2\delta^2(q_1+q_2-\kappa)
d^2q'_1d^2q'_2\delta^2(q'_1+q'_2-\kappa')\eta(q_1,q_2)\eta(q'_1,q'_2)
\]\beq
\frac{\delta^2T_c(\xi)}{\delta\xi(y,q_1,q_2,b)\delta\xi(y',q'_1q'_2,b')}
\Big|_{\xi=0}.
\eeq

Next we use the property proven in ~\cite{CM} that at any values of $\xi$,
due to the equations of motion,
\beq
-\frac{\delta A_c(\xi)}{\delta\xi(y,q_1,q_2,b)}=
\psid_c(y,q_1,b,\xi)\psi_c(y,q_2,b,\xi).
\eeq
This together with (\ref{zero}) immediately gives a simple
formula for the single inclusive cross-section:
\beq
I_1(y,q)=\int d^2b d^2q_1d^2q_2\delta^2(q_1+q_2-q)\eta(q_1,q_2)
\psid_c(y,q_1,b,\xi=0)\psi_c(y,q_2,b,\xi=0).
\label{single}
\eeq
The  formula for the calculation of the double
functional derivative at $\xi=0$ becomes:
\[
\frac{\delta^2T_c(\xi)}{\delta\xi(y,q_1,q_2,b)\delta\xi(y',q'_1q'_2,b')}
\Big|_{\xi=0}=
\frac{1}{2}\Big[\psid_c(y,q_1,b,\xi)\psi_c(y,q_2,b,\xi)\cdot\psid_c(y',q'_1,b',\xi)
\psi_c(y',q'_2,b',\xi)\]\beq+
\frac{\delta}{\delta\xi(y,q_1,q_2,b)}
\Big(\psid_c(y',q'_1,b',\xi)\psi_c(y',q'_2,b',\xi)\Big)
\Big]_{\xi=0}.
\label {deriv2}
\eeq
So the whole problem is reduced to finding $\psi_c$, $\psid_c$ and their
first derivatives in $\xi$ at $\xi=0$.

\subsection{Unitary transformation to new variables and equations of motion}
Considerable simplification can be achieved if, following ~\cite{CM},
we make a unitary transformation of our fields introducing
new fields $\phi_{\pm}$, $\phi_0$ and conjugate fields
$\pi_{\pm}$, $\pi_0$ by
\[
\psi_{\pm}=\phi_{\pm},\ \ \psi_c=-\frac{1}{\sqrt{2}}(\phi_0-\phi_+-\phi_-),
\]
\beq
\psid_{\pm}=\pi_{\pm}+\pi_0,\ \ \psid_c=\sqrt{2}\pi_0.
\eeq
This transforms $L_c$ in to a new Lagrangian
\[
L'_c=\sum_{i=+,-,0}\pi_iQ\phi_i+
\lambda\sum_{i=+,-,0}\Big((K\pi_i)\cdot\phi_i^2+(K\phi_i)\cdot\pi_i^2\Big)
\]\beq
+\lambda\Big(\pi_+\pi_0\cdot K(\phi_0+\phi_+-\phi_-)
+\pi_-\pi_0\cdot K(\phi_0-\phi_++\phi_-))
+\pi_+\pi_-\cdot K(\phi_++\phi_--\phi_0)\Big).
\eeq
The coupling to the nuclei takes the form
\beq
L'_E=g\delta(y)\phi_0+f\delta(y-Y)(\pi_++\pi_-)
\eeq
and the part of action depending on $\xi$ becomes
\beq
-\Delta A_c=\int dyd^2q_1d^2q_2\xi(y,q_1,q_2)
\pi_0(q_1)\Big(\phi_+(q_2)+\phi_-(q_2)-\phi_0(q_2)\Big).
\eeq

Variation of the action with respect to our fields gives  equations of
motion. It can be seen that they are totally identical for $\pm$
fields. So it is sufficient to write them down for
$\phi=\phi_+=\phi_-$, $\phi_0$, $\pi=\pi_+=\pi_-$ and $\pi_0$
They are (at point $y,q$):
\beq
\pi Q+2\lambda K(\pi^2))+2\lambda \phi
K\pi
+\int d^2k\xi(y,q,k)\pi_0(k)=0,
\label{eq1}
\eeq
\beq
\pi_0 Q+\lambda K(\pi^2_0-\pi^2+2\pi_0\pi)+2\lambda\phi_0\cdot K\pi
-\int d^2k\xi(y,q,k)\pi_0(k)+g\delta(y)=0,
\label{eq2}
\eeq
\beq
Q\phi+\lambda K\phi^2+\lambda \pi\cdot K(4\phi-\phi_0)
+\lambda \pi_0\cdot K\phi_0+f\delta(Y-y)=0,
\label{eq3}
\eeq
\beq
Q\phi_0+\lambda K\phi_0^2+2\lambda(\pi_0+\pi)\cdot K\phi_0
+\int d^2k\xi(y,q,k)\Big(2\phi(k)-\phi_0(k)\Big).
\label{eq4}
\eeq

\section{Inclusive cross-sections due to emission from the pomeron}
\subsection{Solution of the equation of motion: fields at $\xi=0$ and
  single inclusive cross-sections}
\
Our first task is to find  the solutions of our equations of motion at
$\xi=0$, which knowledge is sufficient for the single inclusive
cross-section according to (\ref{single}).

If $\xi=0$ then the equations for $\pi$ and $\phi_0$ do not contain driving
terms, so that these fields are identically zero:
\beq
\pi(y,q)=\phi_0(y,q)=0.
\eeq
The two remaining equations decouple:
\beq
\pi_0 Q+\lambda K(\pi_0^2)
+g\delta(y)=0
\eeq
and
\beq
Q\phi+\lambda K\phi^2
+f\delta(Y-y)=0.
\eeq
These are the standard BK equations for the sum of fan diagrams $\chi$
which go to the point $(y,q)$ from the target
\beq
\pi_0(y,q)=\chi(y,q,g)
\eeq
or from the projectile
\beq
\phi(y,q)=\chi(Y-y,q,f)\equiv \tilde{\chi}.
\eeq

Put into the expression (\ref{single}) they give the commonly
used factorized expression for the single inclusive cross-section as
a convolution of two sets of fan diagrams:
\beq
I_1(y,\kappa)=2\int d^2bd^2q_1d^2q_2\delta^2(q_1+q_2-\kappa)\eta(q_1,q_2)
\chi(y,q_1,b)\tilde{\chi}(y,q_2,b)
\label{incpom}
\eeq
or, in the coordinate space, in terms of $Z(y,r,b)$ related to $\chi(y,q,b)$
similarly to (\ref{fromqtor}):
\beq
Z(y,r,b)=2\pi r^2\int \frac{d^2q}{2\pi}e^{iqr}\chi(y,q,b),
\label{fromqtor1}
\eeq
as
\beq
I(y,\kappa)=\frac{8 \alpha_s N_c}{\kappa^2}\int d^2bd^2r Z(y,r,b)\nabla^2 e^{i\kappa r}
\nabla^2\tilde{Z}(y,r,b).
\eeq

\subsection{Derivative of the fields in $\xi$ at $\xi=0$ and double
  inclusive cross-sections}
To set up equations for the derivatives of the fields in $\xi$ at $\xi=0$
we have to differentiate Eqs. (\ref{eq1}) - (\ref{eq4}) in
$\xi(y_1,q_1,q_2,b_1)$
and then put $\xi=0$. The field derivatives and the equations as a whole
will depend on
7 variables: $y,q,b$ and $y_1,q_1,q_2, b_1$, which we shall show
explicitly only when it is necessary. Thus
say $\delta \pi/\delta\xi$ will in fact mean
$\delta\pi(y,q,b)/\delta\xi(y_1,q_1,q_2,b_1)$.
With these notations we get the equations:
\beq
\Big(-\frac{\partial}{\partial
y}+H+\lambda\tilde{\chi}\Big)2K\frac{\delta\pi}{\delta\xi}+
\delta(y-y_1)\delta^2(b-b_1)\delta^2(q-q_1)\chi(q_2)=0,
\label{eq5}
\eeq
\beq
2K\Big(-\frac{\partial}{\partial
y}+H+\lambda\chi\Big)\frac{\delta\pi}{\delta\xi}+
2\lambda\frac{\delta\phi_0}{\delta\xi}K\chi+
2\lambda K\Big(\chi\frac{\delta\pi}{\delta\xi}\Big)
-\delta(y-y_1)\delta^2(b-b_1)\delta^2(q-q_1)\chi(q_2)=0,
\label{eq6}
\eeq
\beq
2K\Big(\frac{\partial}{\partial
y}+H+\lambda\tilde{\chi}\Big)\frac{\delta\phi}{\delta\xi}+
4\lambda\frac{\delta\pi}{\delta\xi}K\tilde{\chi}+
\lambda\chi\cdot  K\frac{\delta\phi_0}{\delta\xi}=0,
\label{eq7}
\eeq
\beq
\Big(\frac{\partial}{\partial
y}+H+\lambda\chi\Big)2K\frac{\delta\phi_0}{\delta\xi}+
+2\delta(y-y_1)\delta^2(b-b_1)\delta^2(q-q_1)\tilde{\chi}(q_2)=0.
\label{eq8}
\eeq

To simplify these multivariable equations we
first note that the $b$ dependence of the derivatives is trivial:
obviously they all are proportional to $\delta^2(b-b_1)$.
So we separate this factor and consider the derivatives at a fixed point
$b$ which need not be shown explicitly. Next we
return to our expression for
the double inclusive cross-section. Obviously it consists of two terms.
One is just the product of
two single inclusive cross-sections
\beq
I_2^{(1)}(y_1,\kappa_1|y_2,\kappa_2)=
I_1(y_1,\kappa_1)I(y_2,\kappa_2).
\label{inclprod}
\eeq

The other term comes from the second term in (\ref{deriv2}) and has the form
\[
I_2^{(2)}(y,\kappa|y',\kappa')=
\int d^2bd^2b'd^2q_1d^2q_2\delta^2(q_1+q_2-\kappa)
d^2q'_1d^2q'_2\delta^2(q'_1+q'_2-\kappa')\eta(q_1,q_2)\eta(q'_1,q'_2)
\]
\[
\frac{\delta}{\delta\xi(y,q_1,q_2,b)}\psid_c(y',q'_1,b',\xi)
\psi_c(y',q'_2,b'\xi)
\Big]_{\xi=0}
\]
\[=
\int d^2q_1d^2q_2\delta^2(q_1+q_2-\kappa)
d^2q'_1d^2q'_2\delta^2(q'_1+q'_2-\kappa')\eta(q_1,q_2)\eta(q'_1,q'_2)
\]
\beq
\Big(\psid_c(y',q'_1,b',\xi=0)
\frac{\delta\psi_c(y',q'_2,b',\xi)}{\delta\xi(y,q_1,q_2,b)}\Big|_{\xi=0}
+\psi_c(y',q'_1,b',\xi=0)
\frac{\delta\psid_c(y',q'_2b',\xi)}{\delta\xi(y,q_1,q_2b)}\Big|_{\xi=0}\Big).
\eeq

Due to factor $\delta^2(b-b')$ contained in the derivatives the double
integrations in $b$ and $b'$ in fact turn into one over the common point
$b$. Taking this into account and suppressing this common argument $b$,
in terms of fields $\phi$ and $\pi$ we find
\[
I_2^{(2)}(y,\kappa|y',\kappa')=
\int d^2b d^2q_1d^2q_2\delta^2(q_1+q_2-\kappa)
d^2q'_1d^2q'_2\delta^2(q'_1+q'_2-\kappa')\eta(q_1,q_2)\eta(q'_1,q'_2)
\]\beq
\Big(\chi(y',q'_1)
\frac{\delta(2\phi(y',q'_2,\xi)-\phi_0(y',q'_2,\xi))}
{\delta\xi(y,q_1,q_2)}\Big|_{\xi=0}
+
2\tilde{\chi}(y',q'_1)
\frac{\delta\pi_0(y',q'_2,\xi)}{\delta\xi(y,q_1,q_2)}\Big|_{\xi=0}\Big),
\label{inclu2}
\eeq
where it is assumed that factor $\delta^2(b-b')$ has been dropped from
the derivatives.
From this formula we can conclude that
we do not need our derivatives in $\xi(y_1,q_1,q_2)$ at all
values of its arguments but rather integrated over $q_1$ and $q_2$
with weight $\delta(q_1+q_2-\kappa)\eta(q_1,q_2)$.
Correspondingly we define (suppressing the argument $b$)
\beq
\Pi_0(y,q, y_1,\kappa)=\int d^2q_1d^2q^2\delta^2(q_1+q_2-\kappa)
\eta(q_1,q_2)
\frac{\delta\pi_0(y,q,\xi)}{\delta\xi(y_1,q_1,q_2)}\Big|_{\xi=0},
\label{pi0}
\eeq
\beq
\Pi(y,q, y_1,\kappa)=\int d^2q_1d^2q^2\delta^2(q_1+q_2-\kappa)
\eta(q_1,q_2)
\frac{\delta\pi(y,q,\xi)}{\delta\xi(y_1,q_1,q_2)}\Big|_{\xi=0},
\label{pi}
\eeq
\beq
\Phi_0(y,q, y_1,\kappa)=\int d^2q_1d^2q^2\delta^2(q_1+q_2-\kappa)
\eta(q_1,q_2)
\frac{\delta\phi_0(y,q,\xi)}{\delta\xi(y_1,q_1,q_2)}\Big|_{\xi=0},
\label{phi0}
\eeq
\beq
\Phi(y,q, y_1,\kappa)=\int d^2q_1d^2q^2\delta^2(q_1+q_2-\kappa)
\eta(q_1,q_2)
\frac{\delta\phi(y,q,\xi)}{\delta\xi(y_1,q_1,q_2)}\Big|_{\xi=0}.
\label{phi}
\eeq
In view of expression (\ref{inclu2}) we also introduce
\beq
\Psi(y,q,y_1,\kappa)=2\Phi(y,q,y_1,\kappa)-\Phi_0(y,q,y_1,\kappa).
\label{psi}
\eeq
In terms of these quantities we find
\[
I_2^{(2)}(y,\kappa|y',\kappa')
\]\beq=
\int
d^2bd^2q'_1d^2q'_2\delta^2(q'_1+q'_2-\kappa')\eta(q'_1,q'_2)
\Big(\chi(y',q'_1)\Psi(y',q'_2,y,\kappa)
+
2\tilde{\chi}(y',q'_1)\Pi_0(y',q'_2,y,\kappa)\Big).
\label{inclu3}
\eeq

Integrating our equations (\ref{eq5}) - (\ref{eq8}) over $q_1$ and $q_2$ with
weight
$\delta^2(q_1+q_2-\kappa)\eta(q_1,q_2)$ we get the following system
\beq
\Big(-\frac{\partial}{\partial y}+H+\lambda\tilde{\chi}\Big)2K\Pi
+\delta(y-y_1)\eta(q,\kappa-q)\chi(y,\kappa-q)=0,
\label{eq9}
\eeq
\beq
2K\Big(-\frac{\partial}{\partial y}+H+\lambda\chi\Big)\Pi_0
+2\lambda\Phi_0\cdot K\chi+2\lambda K(\chi\Pi)
-\delta(y-y_1)\eta(q,\kappa-q)\chi(y,\kappa-q)=0,
\label{eq10}
\eeq
\beq
2K\Big(\frac{\partial}{\partial y}+H+\lambda\tilde{\chi}\Big)\Psi
+8\lambda\Pi\cdot K\tilde{\chi}+2\lambda K(\tilde{\chi}\Phi_0)
-2\delta(y-y_1)\eta(q,\kappa-q)\tilde{\chi}(y,\kappa-q)=0,
\label{eq11}
\eeq
\beq
\Big(\frac{\partial}{\partial y}+H+\lambda\chi\Big)2K\Phi_0
+2\delta(y-y_1)\eta(q,\kappa-q)\tilde{\chi}(y,\kappa-q)=0.
\label{eq12}
\eeq

One observes that the first and last equations determine $\Pi$ and
$\Phi_0$
in terms of the known $\chi$ and $\tilde{\chi}$, after which the second
and third equation allow to find $\Pi_0$ and $\Psi$.

Eqs. (\ref{eq9}) - (\ref{eq12}) are convenient for numerical calculations.
However one
can also formally express the solution in terms of Green function of
the operators $\partial/\partial y+H+\lambda \chi$ and
$\partial/\partial y+H+\tilde{\chi}$ to obtain formulas which can
be compared with ~\cite{CM}

\subsection{Formal solution}
We define the Green functions:
\[
G=\Big(\frac{\partial}{\partial y}+ H+\lambda\chi\Big)^{-1},\ \
G^T=\Big(-\frac{\partial}{\partial y}+ H+\lambda\chi\Big)^{-1},\]\beq
\tilde{G}=\Big(\frac{\partial}{\partial y}+
H+\lambda\tilde{\chi}\Big)^{-1},\ \
\tilde{G}^T=\Big(-\frac{\partial}{\partial y}+
H+\lambda\tilde{\chi}\Big)^{-1}
\eeq
Each Green function is an integral operator in $(y,q)$ space, so that
e.g. for $G$ the kernel is $G(y,q|y',q')$. In our formulas rapidities and
momenta enter asymmetrically, so in many cases we shall suppress the
momenta but leave the rapidities, considering $G_{yy'}$ as an operator in
the momentum space with the kernel $G_{yy'}(q|q')$.

In these notation solution of Eqs. (\ref{eq9}) and (\ref{eq12}) is
immediaite
\beq
\Pi(y)=-\frac{1}{2}K^{-1}\tilde{G}^T_{y,y_1}(\eta\chi)_{y_1},
\label{eqpi}
\eeq
\beq
\Phi_0(y)=-K^{-1}G_{y,y_1}(\eta\tilde{\chi})_{y_1}.
\label{eqphi}
\eeq

We put these solutions into Eq. (\ref{eq10}) multiplied by $(2K)^{-1}$
to obtain
\[
\Big(-\frac{\partial}{\partial y}+H+\lambda\chi\Big)\Pi_0
-\lambda K^{-1}( K\chi(y))
K^{-1}G_{y,y_1}(\eta\tilde{\chi})_{y_1}\]\beq
-\frac{1}{2}\chi\cdot
K^{-1}\tilde{G}^T_{y,y_1}(\eta\chi)_{y_1}
-\delta(y-y_1)\frac{1}{2}K^{-1}(\eta\chi_{y_1})=0.
\label{eq13}
\eeq
Applying operator $G^T$ we get
\beq
\Pi_0=
\lambda\Big[ G^T K^{-1} (K\chi)
K^{-1}G\Big]_{yy_1}(\eta\tilde{\chi})_{y_1}
+\frac{1}{2}\lambda\Big[G^T\chi
K^{-1}\tilde{G}^T\Big]_{y,y_1}(\eta\chi)_{y_1}
+\frac{1}{2}G^T_{yy_1}K^{-1}(\eta\chi)_{y_1}.
\label{eq14}
\eeq

Similar operations with Eq (\ref{eq11}) first give the equation
\beq
\Big(\frac{\partial}{\partial y}+H+\lambda\tilde{\chi}\Big)\Psi
-2\lambda K^{-1}(K\tilde{\chi})
K^{-1}\tilde{G}^T_{y,y_1}(\eta\chi)_{y_1}
-\lambda \tilde{\chi}K^{-1}G_{y,y_1}(\eta\tilde{\chi})_{y_1}
-\delta(y-y_1)K^{-1}\eta\tilde{\chi}_{y_1}=0,
\label{eq15}
\eeq
which after application of operator $\tilde{G}$ gives
\beq
\Psi=
2\lambda \Big[\tilde{G}K^{-1}(K\tilde{\chi})\cdot
K^{-1}\tilde{G}^T\Big]_{y,y_1}(\eta\chi)_{y_1}
+\lambda\Big[\tilde{G}
\tilde{\chi}K^{-1}G\Big]_{y,y_1}(\eta\tilde{\chi})_{y_1}
+\tilde{G}_{yy_1}K^{-1}\eta\tilde{\chi}_{y_1}.
\label{eq16}
\eeq
One observes that $\Psi$ is obtained from $2\Pi_0$ by the substitutions
\beq
G\to \tilde{G}^T,\ \  \tilde{G}\to G^T,\ \
\chi\leftrightarrow\tilde{\chi}.
\label{protarsym}
\eeq
%%%%%%%%%%%%%%%%%%%%%%%%%%%%%%%%%%%%%%%%%%%%%%%%%%%%%%%%%%%%%%%

\begin{figure}
\epsfxsize 3 in
\centerline{\epsfbox{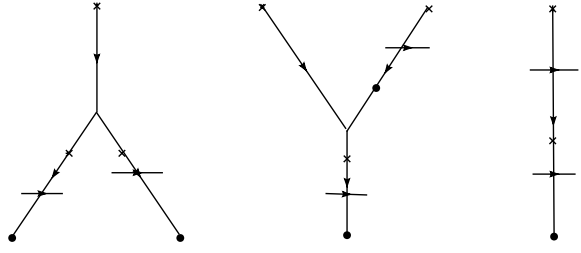}}
\caption{Diagrams corresponding to the non-trivial part $I_2^{(2)}$
of the double emission from pomerons. Crosses mark $\chi$ and $G$,
full circles mark $\tchi$ and $\tilde{G}$.}
\label{fig1}
\end{figure}

Putting the obtained expressions  for the field derivatives
(\ref{eqpi}), (\ref{eqpi}), (\ref{eq14}) and (\ref{eq16})
into (\ref{inclu3})
and using the property (\ref{protarsym}) we obtain the part $I_2^{(2)}$
of the inclusive cross-section as
\[
I_2^{(2)}(y,\kappa|y',\kappa')=\]\[
\int d^2bd^2q'_1d^2q'_2\delta^2(q'_1+q'_2-\kappa')\eta(q'_1,q'_2)
\tchi_{y',q'_1}\Big\{
2\lambda\Big[ G^T K^{-1} (K\chi)
K^{-1}G\Big]_{y',q'_2|y,q_2}(\eta\tilde{\chi})_{y,q_2}\]
\beq
+\lambda\Big[G^T\chi
K^{-1}\tilde{G}^T\Big]_{y',q'_2|,y,q_2}(\eta\chi)_{y,q_2}
+G^T_{y',q'_2|y,q_2}K^{-1}(\eta\chi)_{y,q_2}\Big\}
+\Big(
G\to \tilde{G}^T,\ \  \tilde{G}\to G^T,\ \
\chi\leftrightarrow\tilde{\chi}\Big).
\label{inclu4}
\eeq
%%%%%%%%%%%%%%%%%%%%%%%%%%%%%%%%%%%%%%%%%%%%%%%%%%%%%%%%%%%%%%%%
The obtained expression (\ref{inclu4})
is similar to the one
in the framework of LRFT ~\cite{CM}, except that in our non-local case
all quantities are operators also in the momentum space and that
in various places there appear operators $K$ and $K^{-1}$ acting in
this space. Its diagrammatic illustration is presented in Fig. \ref{fig1}.
In this figure  external lines with crosses (circles) show sums of
fan diagrams propagating towards the target $\chi$ (projectile $\tchi$).
Correspondingly the internal lines marked with crosses
(circles) show the Green functions $G$ ($\tilde{G}$). Horizontal lines
indicate the two observed particles. To the diagrams shown in Fig.1 one
should  add similar ones with the target and projectile interchanged.

From the practical point of view  expression (\ref{inclu4}) is
not very useful due to multiple integrations in $(y,q)$ space.
Numerical solution of evolution equations  (\ref{eq9})-(\ref{eq12})
seems more promising.

\section{Emission from the vertex}
To include  emission from the
triple-pomeron vertex we have to add new parts to the Lagrangian
which describe this emission. In the splitting vertex the pomeron
before the split has always to be cut. The two emerging pomerons may
be both cut and uncut. In accordance with the structure of the
interaction in Eq. (\ref{inter}) the vertex emission part of the
action is to be taken as
\beq
-A_\gamma=\int dy d^2bd^2q_1d^2q_2d^2q_3
L_\gamma +h.c.,
\label{agamma}
\eeq
where
\[
L_{\gamma}=\sqrt{2}\psid_c(q_1)\Big(\gamma_d(q_1|q_2,q_3)\psi_+(q_2)\psi_-(q_3)
-\sqrt{2}\gamma(q_1|q_2,q_3)\psi_c(q_2)(\psi_+(q_3)+\psi_-(q_3))\]\beq+
\gamma_c(q_1|q_2,q_3)\psi_c(q_2)\psi_c(q_3)\Big).
\label{lgamma}
\eeq
The part explicitly shown corresponds to the emission from the
splitting vertex. The part corresponding to the emission from the
merging vertex is
indicated as $h.c.$. The vertex functions $\gamma_d$, $\gamma$ and
$\gamma_c$ describe emissions from the diffractive, single and double
cuts of the triple pomeron vertex respectively. They are real
functions of the three relative momenta of the joining pomerons
and carry  factor $\delta^2(\kappa+q_1-q_2-q_3)$   where
 $\kappa$ is the momentum of the emitted jet.
They are different and their form has been found in
~\cite{bra1} and reproduced in Appendix 2. The fields in (\ref{lgamma}) are
supposed to be taken at
the same rapidity $y$ and transverse point $b$, which dependence is not shown
explicitly.
In the perturbation expansion the new interaction term $A_\gamma$ has
to be taken the number of times which corresponds to the number of
the emissions from the vertex. So for the single inclusive
cross-section we have to take it only once and for the double
inclusive cross-section at most twice.

The single inclusive cross-section corresponding to the emission from the
vertex is obtained from (\ref{lgamma}) just by substituting the fields by
solutions of the equations of motion (\ref{eq1}) - (\ref{eq4}) at $\xi=0$
and integrating over $b$.
For the fields $\psi$ and $\psid$ these solutions are
\beq
\psi_{\pm}=\tilde{\chi},\ \ \psi_c=\sqrt{2}\tilde{\chi},\ \
\psid_{\pm}=\chi,\ \ \psid_c=\sqrt{2}\chi.
\eeq
Thus we find a contribution
\[
I_1^{(\gamma)}(y,\kappa)= 2\int d^2b\prod_{i=1}^3d^2q_i
\Big(\gamma_d(q_1|q_2,q_3)-4\gamma(q_1|q_2,q_3)+
2\gamma_c(q_1|q_2,q_3)\Big)\]
\beq
\Big(\chi(q_1)\tilde{\chi}(q_2)\tilde{\chi}(q_3)+\tilde{\chi}(q_1)
\chi(q_2)\chi(q_3)\Big).
\label{inclga1}
\eeq
The total vertex
\beq
\gamma^{tot}=\gamma_d-4\gamma+2\gamma_c \label{totvert}
\eeq
has a
simple form in the coordinate space.
If we introduce the coordinate vertex $\Gamma(r_1|r_2,r_3)$
acting on non-amputated pomerons according to the relation
\beq
\int d^2q_1d^2q_2d^2q_3\chi(q_1)\gamma(q_1|q_2,q_3)\tchi(q_2)\tchi(q_3)=
\int d^2r_1d^2r_2d^2r_3Z(r_1)\Gamma(r_1|r_2,r_3)\tilde{Z}(r_2)\tilde{Z}(r_3),
\label {eqgaga}
\eeq
where $\chi(q)$ and $Z(r)$ are related by (\ref{fromqtor1}),
then one
finds ~\cite{bra5}
\beq
\Gamma^{tot}(r_1|r_2,r_2)=-\frac{2\alpha_sN_c}{\kappa^2}\nabla^2_1
e^{i\kappa r_1}\nabla^2_1\delta^2(r_2-r_1)\delta^2(r_3-r_1).
\eeq
It corresponds to the expression first obtained in ~\cite{kov} as
a contribution additional to the emission from the pomeron.

Passing to the double inclusive cross-section we first find a
contribution corresponding to the emission from
two verteces, which is obtained by taking
a product of two interactions (\ref{lgamma}) with different external
momenta $\kappa$ and $\kappa'$, at different rapidities
$y$ and $y'$ and transverse points $b$ and $b'$,
substituting in them the fields by the solution
of the equations of motion and integrating over both transverse points
$b$ and $b'$. As a result we obviously find a product of two single
inclusive cross-sections (\ref{inclga1}):
\beq
I_2^{(\gamma\gamma)}(y,\kappa|y',\kappa')=
I_1^{(\gamma)}(y,\kappa)I_1^{(\gamma)}(y',\kappa').
\label{gammaprod}
\eeq

To find the mixed contribution in which one jet is emitted from the
vertex and the other from the pomeron we have to consider the theory
with the interaction term $\Delta A_c$, Eq. (\ref{delta}), and once
differentiate in $\xi$:
\[
I_2^{(\gamma)}(y,\kappa|y'\kappa')=
\int d^2bd^2b'\prod_{i=1}^3d^2q_i d^2q'_1d^2q'_2\delta^2
(q'_1+q'_2-\kappa')\eta(q'_1.q'_2)\]\beq\Big[\frac{\delta}
{\delta\xi(y',q'_1,q'_2,b')}
L_\gamma(y,b,\kappa,q_1,q_2,q_3,\xi)e^{-A_c(\xi)}\Big]_{\xi=0}+
\Big(y\leftrightarrow y',\ \kappa\leftrightarrow \kappa'\Big).
\eeq

Differentiation in $\xi$ will give two terms. One comes from the
differentiation of the exponential. After integrations
over $b'$ and $q'_1$ and $q'_2$ with weight $\eta$ this
differentiation will give  the
single inclusive cross-section $I_1(y',\kappa')$ corresponding to emission
from the pomeron. Factor $L_{\gamma}$ at $\xi=0$ will generate the
single inclusive cross-section (\ref{inclga1}). As a result this part
gives a contribution
\beq
I_1^{(\gamma)}(y,\kappa)I_1(y',\kappa')+
I_1^{(\gamma)}(y',\kappa')I_1(y,\kappa).
\label{mixprod}
\eeq
If we introduce the total single inclusive cross-section
\beq
I^{tot}_1(y,\kappa)=I_1(y,\kappa)+I^{(\gamma)}(y,\kappa)
\eeq
then collecting (\ref{inclprod}), (\ref{mixprod}) and
(\ref{gammaprod}) we find a factorized contribution to the double
inclusive cross-section
\beq
I_2^{fact}(y,\kappa|y'\kappa')=I^{tot}(y,\kappa)I^{tot}(y'\kappa').
\eeq

\begin{figure}
\epsfxsize 3 in
\centerline{\epsfbox{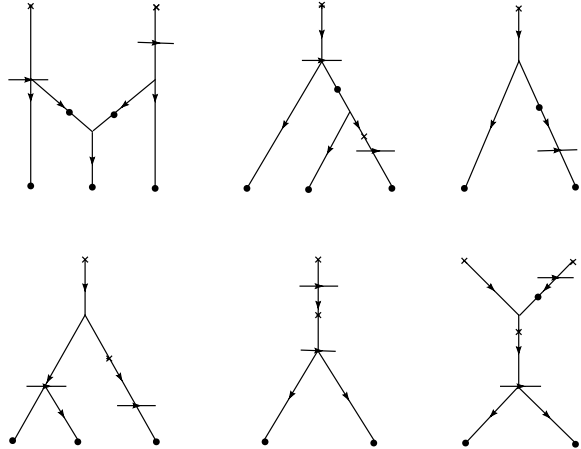}}
\caption{Diagrams for the non-factorized part $I_2^{(\gamma,nf)}$
of the double emission from a pomeron and a vertex. Notations are as in
Fig. \ref{fig1} }
\label{fig2}
\end{figure}

The second part of $I_2^{(\gamma)}$ will come from the differentiation
in $\xi$ of the fields inside $L_{\gamma}$, the exponential factor
giving unity. Differentiation in $\xi$ together with integrations over
$q'_1$ and $q'_2$ with weight $\eta$ will substitute fields in
accordance with Eqs. (\ref{pi0}) - (\ref{psi}). If we define
\beq
D\equiv\int d^2q_1d^2q_2\delta^2(q_1+q_2-\kappa)\eta(q_1,q_2)
\frac{\delta}{\delta\xi(y_1,q_1,q_2}\Big|_{\xi=0}
\eeq
then in terms of fields $\psi$ and $\psid$ we find
\beq
D\psi_c=\frac{1}{\sqrt{2}}\Psi,\ \ D\psi_{\pm}=\Phi,\ \
D\psid_c=\sqrt{2}\Pi_0,\ \ D\psid_{\pm}=\Pi+\Pi_0.
\eeq
We recall that the derivatives are proportional to $\delta^2(b-b')$,
so that the double integration over $b$ and $b'$ turns into a single one.
We then obtain the following expression for the non-factorized part
of the double inclusive cross-section $I^{(\gamma)}_{2}$
\[
I_2^{(\gamma,nf)}(y,\kappa|y'\kappa')=
\int d^2b\prod_{i=1}^3d^2q_i
\Big\{2\gamma^{tot}(q_1|q_1,q_3)
\Pi_0(q_1)\tilde{\chi}(q_2)\tilde{\chi}(q_3)
\]
\[
+2\chi(q_1)\tchi(q_3)\Big[\gamma^{tot}(q_1|q_2,q_3)\Psi(q_2)+
\Big(\gamma_d(q_1|q_2,q_3)-2\gamma(q_1|q_2,q_3)\Big)\Phi_0(q_2)\Big]
\]
\[
+\gamma^{tot}(q_1|q_1,q_3)\Psi(q_1)\chi(q_2)\chi(q_3)
\]\[ +
4\tilde{\chi}(q_1)\chi(q_3)\Big[\gamma^{tot}(q_1|q_2,q_3)\Pi_0(q_2)+
\Big(\gamma_d(q_1|q_2,q_3)-2\gamma(q_1|q_2,q_3)\Big)\Pi(q_2)\Big]
\Big\}\]\beq
+\Big(y\leftrightarrow y',\ \kappa\leftrightarrow \kappa'\Big).
\label{inclga}
\eeq
In this expression it is assumed that the derivative fields
$\Phi_0$, $\Pi$, $\Psi$ and $\Pi_0$, apart from the argument
explicitly shown, depend on their 'own' rapidity $y$ and transverse
point $b$ and also on rapidity $y'$ and external momentum $\kappa'$
which enter in their definitions (\ref{pi0}) - (\ref{psi}).
Due to property (\ref{protarsym}) the last two terms in  (\ref{inclga})
are obtained from the first two ones by interchanging the target and
projectile.
Graphical illustration of $I_2^{(\gamma,nf)}$ is presented in Fig. \ref{fig2},
which shows diagrams corresponding to the six terms in
(\ref{inclga}). The notations are as in Fig. \ref{fig1}. To the diagrams
shown in Fig. \ref{fig2} one has to add the diagrams which are obtained by the
interchange of the target and projectile corresponding to the last two
terms in (\ref{inclga}) and also diagrams from the  the interchange
$y\leftrightarrow y',\ \kappa\leftrightarrow \kappa'$.

\section{Conclusions}
We have derived expressions for the single and double inclusive
cross-sections in nucleus-nucleus collisions in the framework of the
perturbative QCD with interacting BFKL pomerons, in the
quasi-classical approximation (without loops). The cross-sections
include terms with emissions both from the pomerons and from the
triple pomeron vertex. The obtained single inclusive cross-sections
are simple. As expected from the AGK rules they reduce to emissions
from the central pomeron in the convolution of two sets of fan
diagrams connecting it with the projectile and target and from the
two neighboring verteces by which this pomeron splits into fans.
This expected form of the single inclusive cross-sections was used in the
calculations in ~\cite{bra4}.

In contrast, the expressions for the double inclusive cross-sections are
much more complicated. A  part of it is just the product of two
single inclusive cross-sections corresponding to independent emissions
of two jets from different points in the nuclear overlap transverse
space. The other part however includes all sorts of rescattering
corrections and needs summation of diagrams of a structure different
from fans. It also includes emissions from the triple pomeron vertex
having different forms for different  cuts passing
through the vertex. We have set up  evolution equations which
allow to find  the rescattering corrections to the double inclusive
cross-sections. Different forms of emission from the vertex have been
found in our paper ~\cite{bra5}. So in principle the theoretical
basis for the calculation of the double inclusive cross-section is
completed. However  practical computations along these lines seem to
be quite complicated and apparently present a separate (and
formidable) task.

\section{Acknowledgements}
This work has been supported by grants of the Ministry of Education
of Russia RNP 2.1.1.112 and RFFI of Russia 06-02-16115-a.

\section{ Appendix 1. The unitarity equation with cut pomerons}

The derivation with BFKL pomerons follows that in ~\cite{CM} for local
pomerons.
The starting point is the  equation which tells that the sum of all
Green functions in the cut pomeron theory is equal to the sum of all
Green function in the original theory, provided the couplings of the
fields to themselves and  external sources are chosen in accordance
with the AGK rules.
It is written as
\beq
\sum_{n,m}\frac{g^n}{n!}G^{(n,m)}\frac{f^m}{m!}=
{\sum_{n_i,m_j}}'\frac{ig^n(\sqrt{2})^{n_c-1}}{n_+!n_-!n_c!}
G^{(n_i,m_j)}\frac{if^m(\sqrt{2})^{m_c-1}}{m_+!m_-!m_c!}.
\label{sum1}
\eeq
Here $G^{(n,m)}$ are the Green function in the original theory with
$n$ incoming and $m$ outgoing pomerons and $G^{(n_i,m_j)}$ are the
Green function in the cut pomeron theory, where
$i,j=+,-,c$, $n_i$ is the number of incoming pomerons of sort $i$,
$n=n_++n_-+n_c$ and  $m_j$ is the number of outgoing pomerons of sort $j$,
$m=m_++m_-+m_c$. Symbol $\sum'$ means that one
should drop terms with $n_\pm=n_c=0$ and $m_\pm=m_c=0$ which are
absent in the cut theory. The Green functions are assumed to be
operators acting in the rapidity-transverse momentum space,
convoluted with the external sources given by functions $g$ and $f$.

One then uses obvious properties
\beq
G^{(n_{\pm}=n,n_{\mp}=n_c=0;m_{\pm}=m,m_{\mp}=m_c=0)}=G^{(n,m)},
\label{prop1}
\eeq
\beq
G^{(n_{\pm}=n,n_{\mp}=n_c=0;m_{\mp}=m,m_{\pm}=m_c=0)}=0
\label{prop2}
\eeq
to consider the sum (\ref{sum1}) without restrictions on the
summations over $n_i$ and $m_j$ and multiplied by 2.
One has
\[
\sum_{n_i,m_j}\frac{ig^n(\sqrt{2})^{n_c}}{n_+!n_-!n_c!}
G^{(n_i,m_j)}\frac{if^m(\sqrt{2})^{m_c}}{m_+!m_-!m_c!}
\]\beq
=2\sum_{n,m>0}
\sum_{n_i,m_j}
2\frac{ig^n(\sqrt{2})^{n_c-1}}{n_+!n_-!n_c!}
G^{(n_i,m_j)}\frac{if^m(\sqrt{2})^{m_c-1}}{m_+!m_-!m_c!}
+2\sum_{n,m>0}\frac{g^n}{n!}G^{n,m)}\frac{f^m}{m!}+1.
\label{sum2}
\eeq
where in the first sum of the second equality $n=n_++n_-+n_c$
and $m=m_++m_-+m_c$.
However according to (\ref{sum1}) the first term is exactly equal to
the second with the opposite sign. So we are left with
\beq
\sum_{n_i,m_j}\frac{ig^n(\sqrt{2})^{n_c}}{n_+!n_-!n_c!}
G^{(n_i,m_j)}\frac{if^m(\sqrt{2})^{m_c}}{m_+!m_-!m_c!}
=1.
\label{sum3}
\eeq
The sum in (\ref{sum3}) is just the total $S$ matrix in the cut pomeron
theory, so that the equality (\ref{sum3})
means that action $A_c(\xi=0)=0$.

In the theory with the action $A_c(\xi)$ the amplitude $T_c(\xi)$ is
\beq
T_c(\xi)=
\frac{1}{2}{\sum_{n_i,m_j}}'\frac{g^n(\sqrt{2})^{n_c}}{n_+!n_-!n_c!}
G^{(n_i,m_j)}_c(\xi)\frac{f^m(\sqrt{2})^{m_c}}{m_+!m_-!m_c!}.
\label{sum4}
\eeq
Adding and subtracting terms with $n_+=n_c=0$ and $m_-=m_c=0$
and using properties (\ref{prop1}) and (\ref{prop2})
we find
\beq
T_c(\xi)=\frac{1}{2}\sum_{n_i,m_j}
\frac{g^n(\sqrt{2})^{n_c}}{n_+!n_-!n_c!}
G^{(n_i,m_j)}_c(\xi)\frac{f^m(\sqrt{2})^{m_c}}{m_+!m_-!m_c!}
-\sum_{m,n}\frac{g^n}{n!}G^{n,m)}\frac{f^m}{m!}+\frac{1}{2},
\label{sum5}
\eeq
where again in the first sum $n=n_++n_-+n_c$
and $m=m_++m_-+m_c$.
The first sum is just one half of $e^{-A_c(\xi)}$
while the second is $e^{-A}$ in the original theory, so that
Eq. (\ref{sum5}) leads to Eq. (\ref{tcxi}).

\def\bh{{\bf h}}

\section{Appendix 2. Cut vertex functions $\gamma_d$, $\gamma$ and $\gamma_c$}

The three cut vertex functions $\gamma_d$, $\gamma$ and $\gamma_c$ which enter
(\ref{lgamma}) can be most conveniently written in the coordinate space
as functions $\Gamma_d$, $\Gamma$ and $\Gamma_c$ acting on non-amputated
pomeron functions according to (\ref{eqgaga}).
To simplify notations we introduce a vector
\beq
\bh(r_1,r_2)=\frac{{\bf r}_1}{r_1^2}-\frac{{\bf r}_1+{\bf r}_2}{(r_1+r_2)^2}.
\eeq
From the results obtained in ~\cite{bra5}) we find
\[
\Gamma_d(r_1|r_2,r_3)=
\frac{2\alpha_s N_c}{(2\pi)^2}\nabla_1^4\]\[
\Big\{e^{i\kappa (r_2-r_3)}\bh(r_2,r_1)\bh(r_3,r_1)
-e^{i\kappa (r_2-r_3+r_1)}\bh(r_2,r_1)\bh(r_3,-r_1)
\]\[
+4i\pi e^{i\kappa r_2}\Big(1-e^{i\kappa r_1}\Big)
\frac{{\bf\kappa}}{\kappa^2}\bh(r_2,r_1)\delta^2(r_3-r_1)
+\frac{4\pi^2}{\kappa^2}(1-e^{i\kappa r_1})\delta^2(r_2-r_1)
\delta^2(r_3-r_1)
\]\beq
-\frac{4\pi^2}{\kappa^2}\nabla^{-2}e^{i\kappa r_1}\nabla_1^2
\delta^2(r_2-r_1\delta^2(r_3-r_1))\Big\},
\eeq
%%%%%%%%%%%%%%%%%%%%%%%%%%%%%%%%%%%%
%%%%%%%%%%%%%%%%%%%%%%%%%%%%%%%%%%%%
\[
\Gamma(r_1|r_2,r_3)=\frac{\alpha_s N_c}{2(2\pi)^2}\nabla_1^4
\]\[
\Big\{e^{i\kappa (r_2-r_3)}\bh(r_2,r_1)\bh(r_3,r_1)
-2e^{i\kappa (r_2-r_3+r_1)}\bh(r_2,r_1)\bh(r_3,-r_1)
\]\[
-6i\pi ie^{i\kappa r_2}
\frac{{\bf\kappa}}{\kappa^2}\bh(r_2,r_1)\delta^2(r_3-r_2)
-3i\pi e^{i\kappa r_2}\Big(1-e^{i\kappa r_1}\Big)
\frac{{\bf\kappa}}{\kappa^2}\bh(r_2,r_1)\delta^2(r_3-r_1-r_2)
\]\[
-3e^{i\kappa r_3}\bh(r_2,r_1)\bh(r_2+r_3,r_1)
+2i\pi e^{i\kappa r_2}\Big(1-2e^{i\kappa r_1}\Big)
\frac{{\bf\kappa}}{\kappa^2}\bh(r_2,r_1)\delta^2(r_3-r_1)
\]\beq
+\frac{4\pi^2}{\kappa^2}(1+e^{i\kappa r_1})
\delta^2(r_2-r_1)\delta^2(r_3-r_1)-
\frac{16\pi^2}{\kappa^2}\nabla^{-2}e^{i\kappa r_1}\nabla_1^2
\delta^2(r_2-r_1\delta^2(r_3-r_1))\Big\},
\eeq
%%%%%%%%%%%%%%%%%%%%%%%%%%%%%%%%%%%%%%%%%%%%%%%%
%%%%%%%%%%%%%%%%%%%%%%%%%%%%%%%%%%%%%%%%%%%%%%%%%%
\[
\Gamma_c(r_1|r_2,r_3)=\frac{2\alpha_s N_c}{(2\pi)^2}\nabla_1^4
\]\[
\Big\{-e^{i\kappa (r_2-r_3+r_1)}\bh(r_2,r_1)\bh(r_3,-r_1)
-3e^{i\kappa r_3}\bh(r_2,r_1)\bh(r_2+r_3,r_1)
\]\[
-3i\pi e^{i\kappa r_2}\Big(1-e^{i\kappa r_1}\Big)
\frac{{\bf\kappa}}{\kappa^2}\bh(r_2,r_1)\delta(r_3-r_1-r_2)
+2i\pi e^{i\kappa r_2}
\frac{{\bf\kappa}}{\kappa^2}\bh(r_2,r_1)\delta(r_3-r_2)
\]\[
+2i\pi e^{i\kappa r_2}
\frac{{\bf\kappa}}{\kappa^2}\bh(r_2,r_1)\delta(r_3-r_1)
-16\pi^2\nabla_1^{-2}e^{i\kappa r_1}\delta^2(r_2-r_1\delta^2(r_3-r_1)
\]\beq
-\frac{8\pi^2}{\kappa^2}\nabla^{-2}e^{i\kappa r_1}\nabla_1^2
\delta^2(r_2-r_1\delta^2(r_3-r_1)\Big\}.
\eeq
Due to symmetry properties of the pomerons, these expressions have to
be symmetrized in $r_2$ and $r_3$. Also all exponentials have to be
substituted by their real and imaginary parts:
\beq
e^{ikr}\to \cos kr,\ \ ie^{ikr}\to -\sin kr,
\eeq
which makes the cut vertexes real.

%%%%%%%%%%%%%%%%%%%%%%%%%%%%%%%%%%%%%%%%%%%%%%%%%%%%%%%%%%%%%%%%
%%%%%%%%%%%%%%%%%%%%%%%%%%%%%%%%%%%%%%%%%%%%%%%%%%%%%%%%%%%%%%%%

%%%%%%%%%%%%%%%%%%%%%%%%%%%%%%%%%%%%%%%%%%%%%%%%%%%%%%%%%%%%%%%%%%%%%%%%%%%%

\end{document}